\documentclass[aps,prl,reprint,groupedaddress]{revtex4-1}
\pdfoutput=1
\usepackage[T1]{fontenc}
\usepackage{graphicx}
\usepackage{amssymb,amsfonts,amsmath}

\newcommand  {\hb}	 {\hat{\textbf{b}}} 
\newcommand  {\hp} {\hat{\textbf{p}}} 
\newcommand  {\hs} {\hat{\textbf{s}}}
\newcommand  {\vecr} {\textbf{r}}  
\newcommand  {\eij} {\epsilon_{ij}}
\newcommand  {\apa} {a_{\text{p}}} 
\newcommand  {\aap} {a_{\text{ap}}} 
\newcommand  {\xf} {x_{\text{f}}} 
\newcommand  {\mo} {m_0} 
\newcommand  {\wo} {w_0} 
\newcommand  {\Emin}{E_{\text{min}}} 
\newcommand  {\Tm}      {T_{\text{m}}}

\newcommand  {\e}         {\text{e}}
\newcommand  {\Pacc}	{p_{\text{acc}}} 
\newcommand  {\kB}      {k_{\text{B}}}

\begin{document}

%\preprint{}

\title{Aggregate geometry in amyloid fibril nucleation}


\author{Anders Irb\"ack, Sigur\dh ur \AE. J\'onsson, Niels Linnemann,\\ Bj\"orn Linse and
Stefan Wallin}
%\thanks{}
\affiliation{Computational Biology and Biological Physics,
Department of Astronomy and Theoretical Physics, Lund University,
S\"olvegatan 14A, SE-223 62 Lund, Sweden}


\date{\today}

\begin{abstract}
We present and study a minimal structure-based model for the self-assembly of 
peptides into ordered $\beta$-sheet-rich fibrils. The peptides are represented 
by unit-length sticks on a cubic lattice and interact by hydrogen bonding
and hydrophobicity forces. By Monte Carlo simulations with $>$$10^5$
peptides, we show that fibril formation occurs with sigmoidal kinetics in the
model. To determine the mechanism of fibril nucleation, we compute the 
joint distribution in length and width of the aggregates at equilibrium, using 
an efficient cluster move and flat-histogram techniques. This analysis, based 
on simulations with 256 peptides in which aggregates form and dissolve 
reversibly, shows that the main free-energy barriers that a nascent fibril 
has to overcome are associated with changes in width. 
\end{abstract}

\pacs{87.14.em,87.15.A-}

%\keywords{}

\maketitle

Many proteins and peptides share the ability to self-assemble into 
amyloid fibrils, aggregates with a cross-$\beta$ structure and remarkable 
mechanical properties, that are associated with a range of disorders 
as well as with functional roles~\cite{Chiti:06,Knowles:11}. 
The formation of amyloid fibrils, usually monitored by thioflavin T (ThT) fluorescence, 
is known to occur with reproducible sigmoidal kinetics~\cite{Hellstrand:10}, 
indicating a nucleation-dependent process. A powerful method for interpreting
the experimental kinetic profiles is by means of rate equations~\cite{Knowles:09}.   
This approach can reveal some general properties of intermediate species 
participating in the growth process. It has 
proven useful for some related self-assembly phenomena 
as well,  such as hemoglobin S aggregation~\cite{Ferrone:85} and microtubule
assembly~\cite{Flyvbjerg:96}. Another 
method to elucidate 
the mechanisms of amyloid formation is by phase 
equilibria analysis~\cite{Schmit:11}. 

By coarse-grained structure-based approaches~\cite{Wu:11}, additional 
insights have been gained into the nucleation of amyloid 
fibrils~\cite{Pellarin:07,Auer:08,LiMS:08,Zhang:09,Auer:10,Bieler:12}. Fibrillation 
pathways involve, however, a host of different aggregated species of 
widely varying size, and studying the competition among these species 
without restrictive assumptions represents a challenge even in 
coarse-grained models.  

In this Letter, we introduce a minimal structure-based model that   
describes amyloid fibril formation in terms of 
physically inspired peptide-peptide interactions and yet allows for representative 
sampling of the model state space for relatively large systems.
Using flat-histogram methods~\cite{Berg:91,Wang:01a} 
and an efficient cluster move resembling the Swendsen-Wang algorithm 
for spin systems~\cite{Swendsen:87}, we determine equilibrium distributions 
in size and shape of the aggregated structures, in order to elucidate the 
free-energy landscape that a nascent fibril has to navigate. 

We consider $N$ identical peptides, represented by unit-length sticks
on a periodic cubic lattice with dimensions $L^3$.  We assume that the 
internal dynamics of a peptide are fast compared to the timescales for 
fibril formation, and therefore can be averaged out.  

\begin{figure}[t]
\includegraphics[width=8cm]{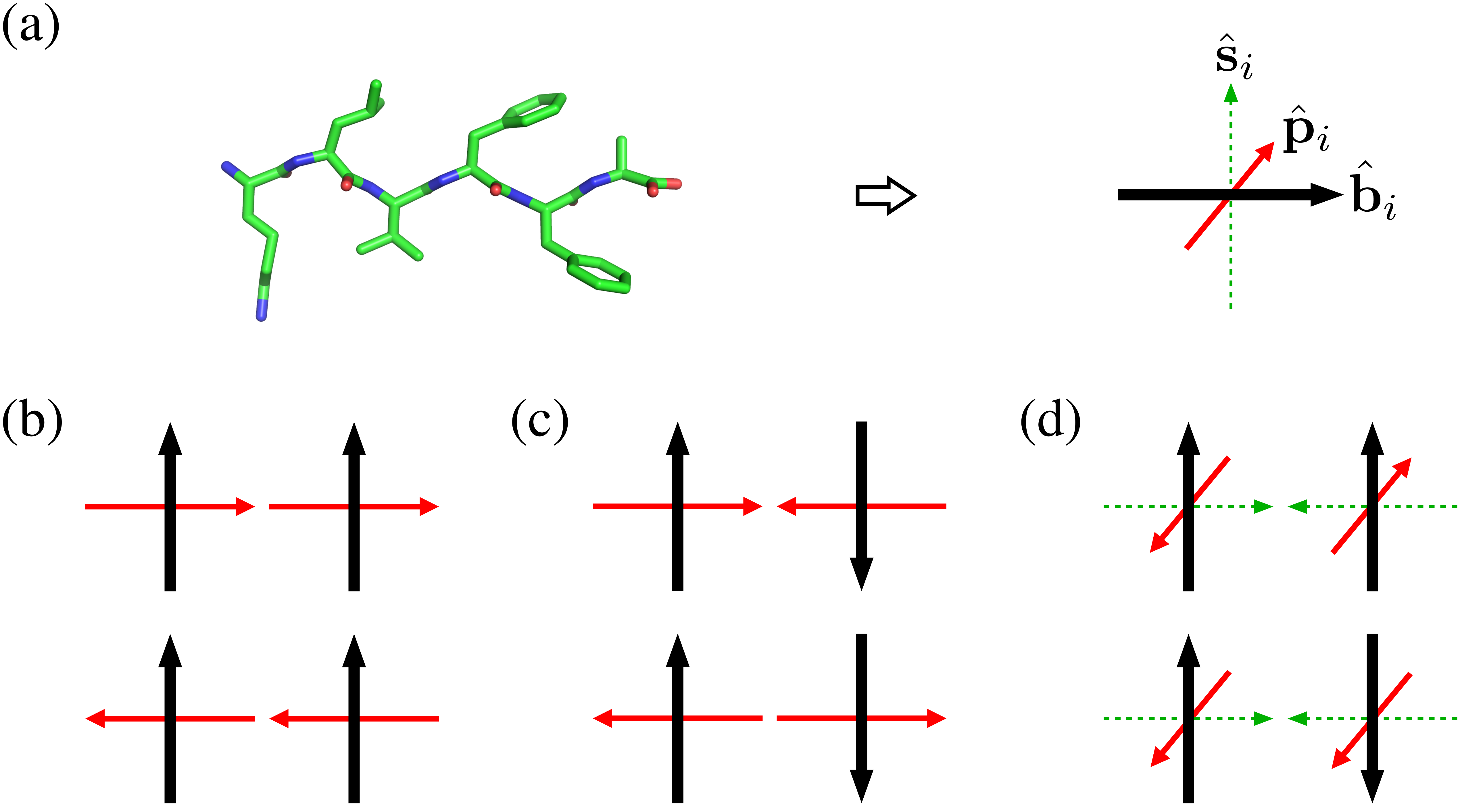}
\caption{Schematic illustration of the geometry and interactions of the model.
(a) The orientation of a peptide is defined by the backbone vector $\hb_i$ (thick line) 
and the hydrogen-bond direction $\hp_i$ (thin line). The side-chain direction $\hs_i$ 
(dots) is given by $\hs_i=\hb_i\times\hp_i$. (b) Parallel $\beta$-structure.
(c) Antiparallel $\beta$-structure. (d) Hydrophobic attraction. The $\beta$-structure 
definitions are such that a $\beta$-sheet has one hydrophobic 
and one polar side.}
\label{fig:model}
\end{figure}

Each peptide $i$ is centered at a lattice site, $\vecr_i$, and two 
peptides cannot simultaneously occupy the same site.
Associated with each peptide are two unit vectors $\hb_i$ and 
$\hp_i$ that can point in any of the six lattice directions 
(Fig.~\ref{fig:model}a); $\hb_i$ represents the N-to-C backbone 
orientation, whereas $\pm\hp_i$ are the directions in which 
hydrogen bonds can form. The vectors $\hb_i$ 
and $\hp_i$ are perpendicular, leaving a total of 24 possible 
orientations of a peptide. The vectors  $\hs_i=\hb_i\times\hp_i$ and 
$-\hs_i$ represent side-chain directions. The $+\hs_i$ and $-\hs_i$ sides 
of a peptide are assumed to have different interaction properties and are 
referred to as hydrophobic and polar, respectively. 

The energy function describing the interactions between the peptides 
is assumed pairwise additive, $E=\sum_{i<j}\eij$, where $\eij\le0$. The 
pair potential $\eij$ is nonzero only if (i) peptides $i$ and $j$ are nearest 
neighbors on the lattice, and (ii) $\hb_i$ and $\hb_j$ are perpendicular 
to $\vecr_{ij}=\vecr_j-\vecr_i$ and aligned either parallel or antiparallel to each other. 
When these conditions are met, we set $\eij=-1$ except in the three cases illustrated
in Fig.~\ref{fig:model}. The first two cases correspond to parallel (Fig.~\ref{fig:model}b) 
and antiparallel (Fig.~\ref{fig:model}c) $\beta$-structure, respectively, and 
the third (Fig.~\ref{fig:model}d) to hydrophobic side-chain attraction. The 
corresponding interaction energies are given by 
\begin{equation}     
\label{eq:energies}
\eij=
\begin{cases}
  -(1+\apa) & \text{parallel $\beta$-structure} \\
  -(1+\aap) & \text{antiparallel $\beta$-structure} \\
  -(1+b) & \text{hydrophobic attraction}
\end{cases}
\end{equation}    
The hydrophobic 
attraction is included because of evidence suggesting that a 
pairwise (steric zipper) $\beta$-sheet organization is a common 
architecture for the core of amyloid fibrils~\cite{Sawaya:07}. 
The $b$ parameter must not be too large, in order for 
extended $\beta$-sheets to form. Because the 
$\beta$-sheets often are parallel in amyloid fibrils, 
we take $\apa>\aap$, but the model can 
also be studied for $\aap>\apa$. In what follows, for simplicity, we 
stick to a single parameter set, namely  $\apa=5$, $\aap=3$ 
and $b=1$. With this choice, the parallel $\beta$-strand
organization dominates, but the suppression of antiparallel 
strand pairs is not prohibitively strong.

We simulate the thermodynamics of this model using single-peptide as well as cluster 
moves. A cluster update makes it possible for aggregates to move 
without having to be first dissolved and then reassembled. To be able 
to also split and merge aggregates, we follow a stochastic 
Swendsen-Wang-type cluster construction 
procedure~\cite{Swendsen:87}. The construction is recursive and begins
by picking a random first cluster member, $i$. Then, all peptides $j$
interacting with peptide $i$ ($\eij<0$) are identified, and added to the 
cluster with probability $p_{ij}=1-\e^{\beta \eij}$, where $\beta=1/\kB T$  
is inverse temperature. This step is iterated until no
cluster member has any unchecked interaction partner. Finally, the 
resulting cluster is subject to a trial rigid-body translation or rotation,
drawn from a symmetric distribution, which is accepted whenever it does not 
cause any steric clashes. It can be verified that this algorithm
fulfills detailed balance with respect to the canonical 
ensemble $p_\nu\propto\e^{-\beta E_\nu}$.       
 
To further enhance the sampling, we employ generalized-ensemble 
methods~\cite{Wang:01a,Berg:91}, along with reweighting 
techniques~\cite{Ferrenberg:89}. After estimating the density of 
states, $g(E)$, by the Wang-Landau method~\cite{Wang:01a}, 
we simulate the ensemble $p_\nu\propto1/g(E_\nu)$~\cite{Berg:91},  
where the distribution of $E$ is flat. This approach was recently used  
for atomic-level aggregation simulations~\cite{Jonsson:11} 
and is useful for the present system as well, which displays phase
coexistence at the fibrillation temperature, $\Tm$ 
(see below). Our simulations sample a limited energy 
range, $\Emin<E\le0$. The cutoff $\Emin$ is needed to prevent the 
formation of artificial cyclic aggregates, which otherwise may occur 
due to the periodic boundary conditions, but is sufficiently low 
to permit studies of temperatures in the fibrillar phase. 

The above cluster update can be adapted for the 
generalized-ensemble simulations
by adding an accept/reject step, with acceptance 
probability $\Pacc(\nu\to\nu^\prime)=
\min[1,g(E_\nu)\e^{-\beta E_\nu}/g(E_{\nu^\prime})\e^{-\beta E_{\nu^\prime}}]$.
Here, $\beta$ changes its meaning to become a tunable algorithm 
parameter. We did not fine-tune $\beta$, but expect the optimal $\beta$ 
to be in the neighborhood of $\beta_{\text{m}}=1/\kB\Tm$, as supported by 
preliminary runs.

Using these methods, we studied the thermodynamics of the 
model for several different system sizes. Here, we focus 
on the results obtained for $N=256$ and $L=64$, corresponding 
to a peptide concentration of $\rho\approx 10^{-3}$ per unit 
volume. This system size would have been very time-consuming 
to study with standard Monte Carlo methods. 

\begin{figure}[t]
\includegraphics[width=7.5cm]{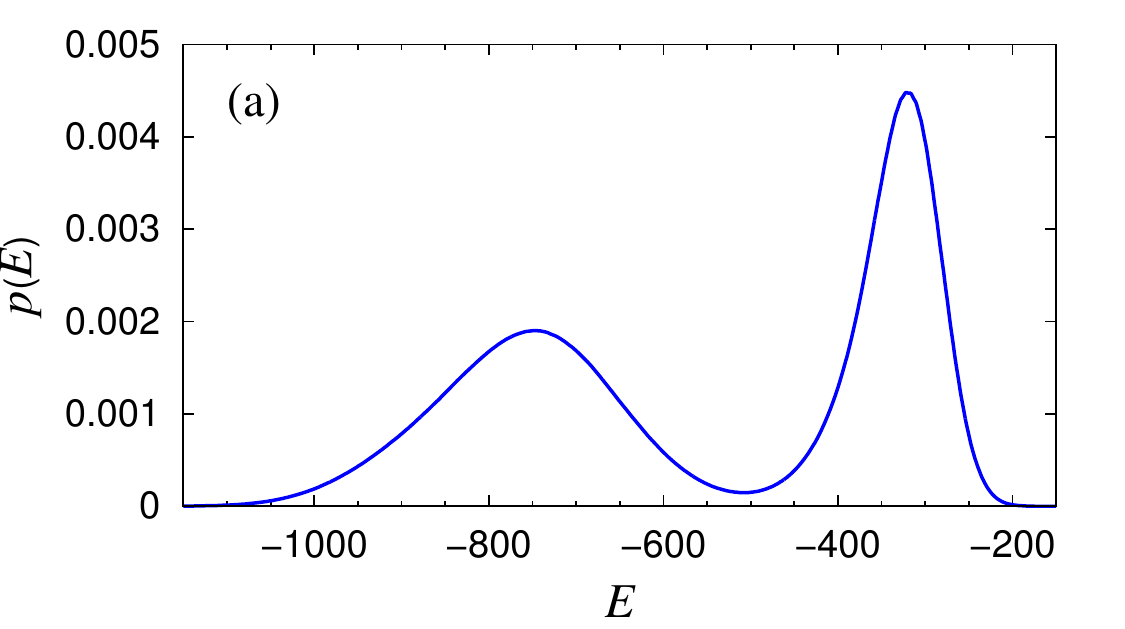}
\includegraphics[width=7.5cm]{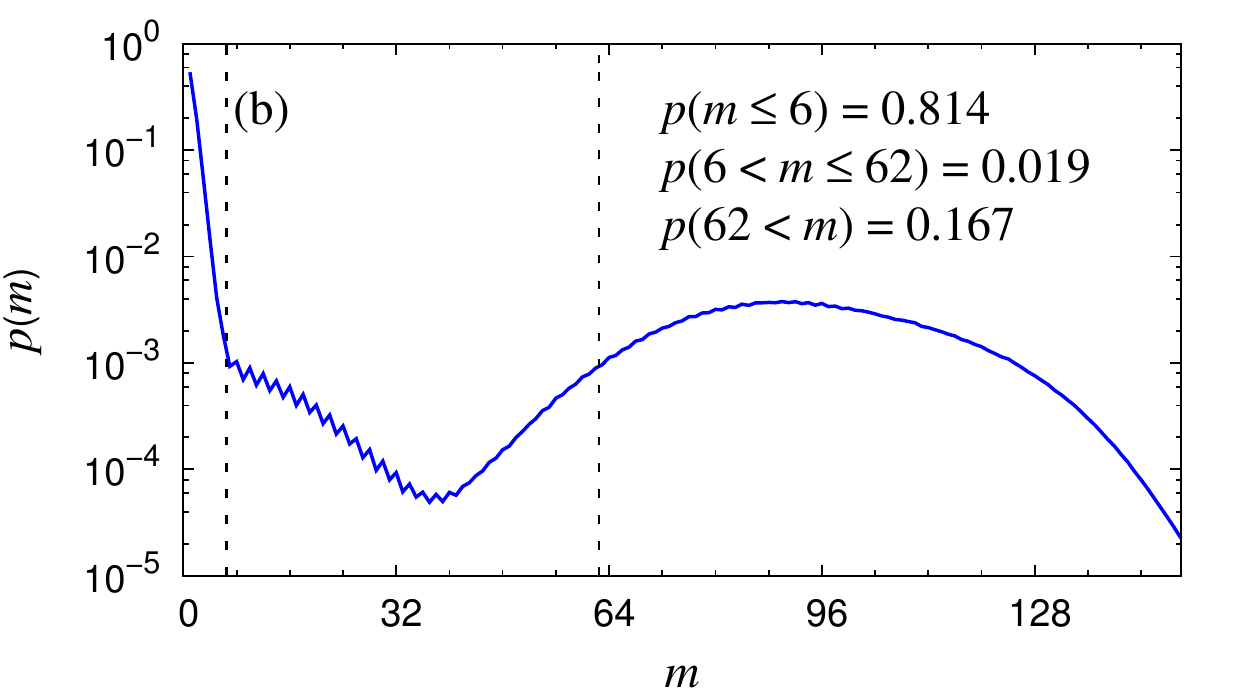}
\caption{Overall thermodynamic properties of the $N=256$, $L=64$ 
system at $\Tm\approx 0.6714$. (a) Energy distribution. 
The sampled range is $-1300<E\le 0$. Smoothing 
was applied to remove short-scale irregularities in $g(E)$. 
(b) Mass fraction of aggregates with mass $m$, $p(m)$, against $m$.  
Summed probabilities for three regions in $m$ are indicated. 
Statistical errors on both $p(E)$ and $p(m)$ are comparable to 
the line width.}
\label{fig:distr}
\end{figure}

In our simulations, two distinct major phases occur: a high-energy 
phase dominated by small aggregates and a low-energy phase 
where large fibril-like aggregates are present. As displayed in
Fig.~\ref{fig:distr}a, at the midpoint temperature, $\Tm\approx0.6714$, 
the energy distribution is bimodal, showing that the two phases coexist.
Fig.~\ref{fig:distr}b shows the aggregate mass distribution, $p(m)$, at $\Tm$, 
which gives the probability for a peptide to be part of an aggregate with 
$m$ peptides ($m=1$ corresponds to free monomers). Like the energy 
distribution, $p(m)$ is bimodal. The mass fractions of aggregates with 
$m\le6$, $6<m\le62$ and $m>62$ are 81.4\%, 1.9\% and 16.7\%, respectively. 
Small aggregates are present in both phases
and constitute a large fraction of the total mass at $\Tm$
(see also Supplemental Material, Fig.~S1~\cite{SUPP}). 

At first glance, the bimodality of $p(m)$ may seem to indicate that 
fibril nucleation occurs when a critical mass is reached. However, 
this picture is geometrically incomplete,  
because the species involved are neither strictly 
one-dimensional nor sharing one common shape, such as spherical.  
A simple but useful way to extend the analysis is via the inertia 
tensor. As measures of the length and width of an aggregate, 
we define $l=\sqrt{12\lambda_1+1}$ and $w=\sqrt{12\lambda_2+1}$, 
where $\lambda_1\ge\lambda_2$ are eigenvalues of the 
inertia tensor. 
In our model, there is no interaction 
between longitudinally aligned peptides to support growth in a 
third dimension. With these definitions, for a %perfectly 
rectangular aggregate, $l$ and $w$ are the numbers of peptide layers 
in the two directions. 

\begin{figure}[t]
\includegraphics[width=8.4cm]{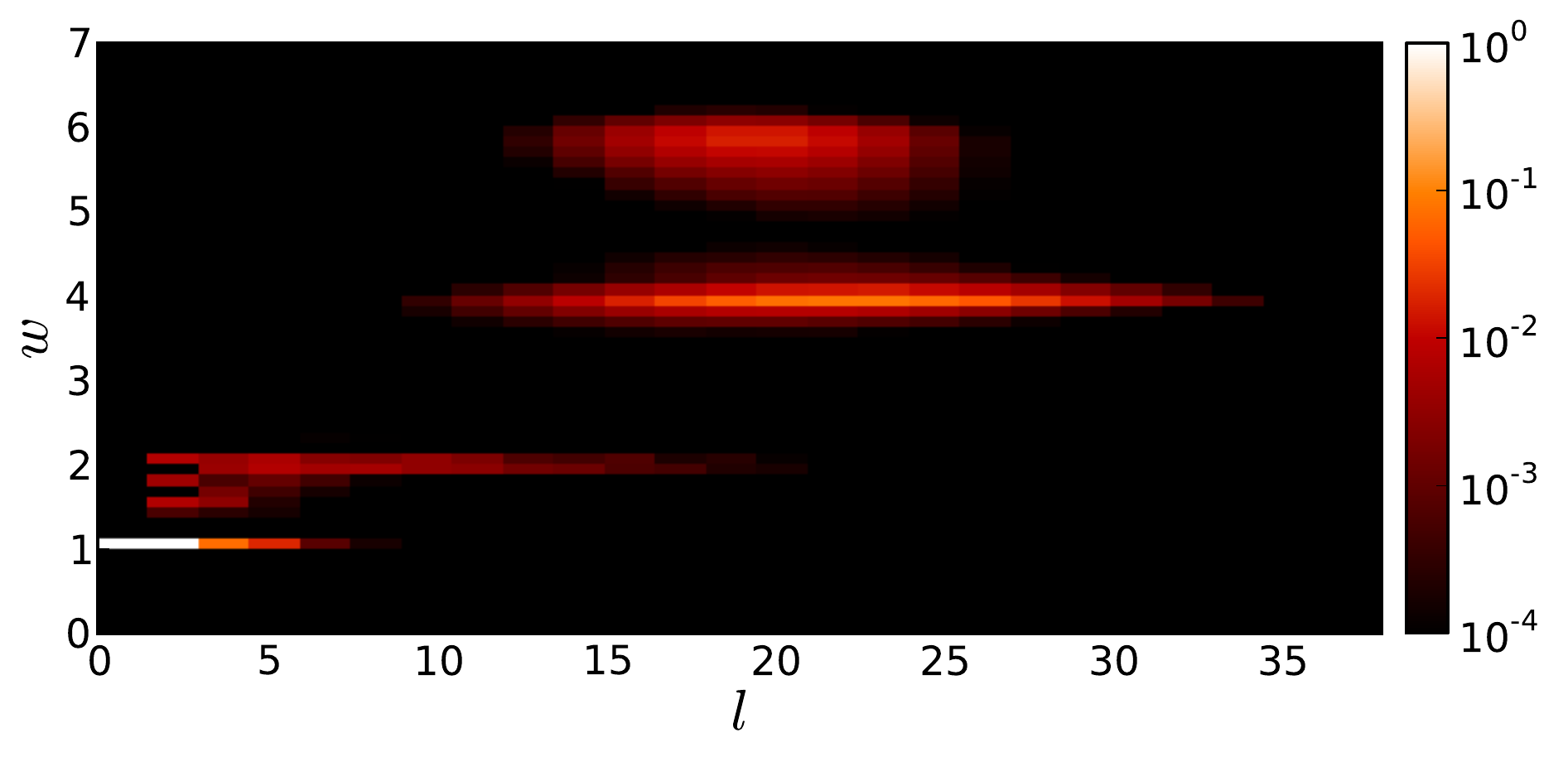}
\caption{Mass fraction of aggregates with length $l$ and width $w$,  
$p(l,w)$, at $\Tm$ for $N=256$ and $L=64$.\label{fig:lw}} 
\end{figure}

Fig.~\ref{fig:lw} shows the probability $p(l,w)$ for a peptide to be part 
of an aggregate with length $l$ and width $w$, at $\Tm$. 
Consistent with Fig.~\ref{fig:distr}b, 
$p(l,w)$ is highest for aggregates with $l$ small
and $w\approx1$. Among larger aggregates, a clear preference can be 
seen for even over odd values of $w$, reflecting a pairwise $\beta$-sheet 
organization, although aggregates with six or more layers 
are severely constrained by finite-size effects.
A second trend is that single-layer aggregates are 
shorter than two-layer ones, which in turn are shorter than those 
with four layers. We expect this trend to persist beyond the 
four-layer level if the system is sufficiently large. These overall features 
of  $p(l,w)$ are likely to be quite robust, although the locations
of the different maxima depend on both $T$ and $\rho$.  
  
The shape of $p(l,w)$ has implications for how fibrils nucleate and grow
in the model. It suggests that the main free-energy barriers faced
by a growing aggregate are associated with changes in width, and it 
must increase in width to be able to grow. 

Having examined the thermodynamics of the model, we now turn to 
the aggregation kinetics, studied using constant-temperature 
Monte Carlo dynamics. Because of evidence that amyloid growth 
occurs by monomer addition~\cite{Collins:04}, we here use single-peptide 
moves only. The simulations start from random initial conditions and 
the temperature is $T=0.66$. 

\begin{figure}[t]
\includegraphics[width=7.5cm]{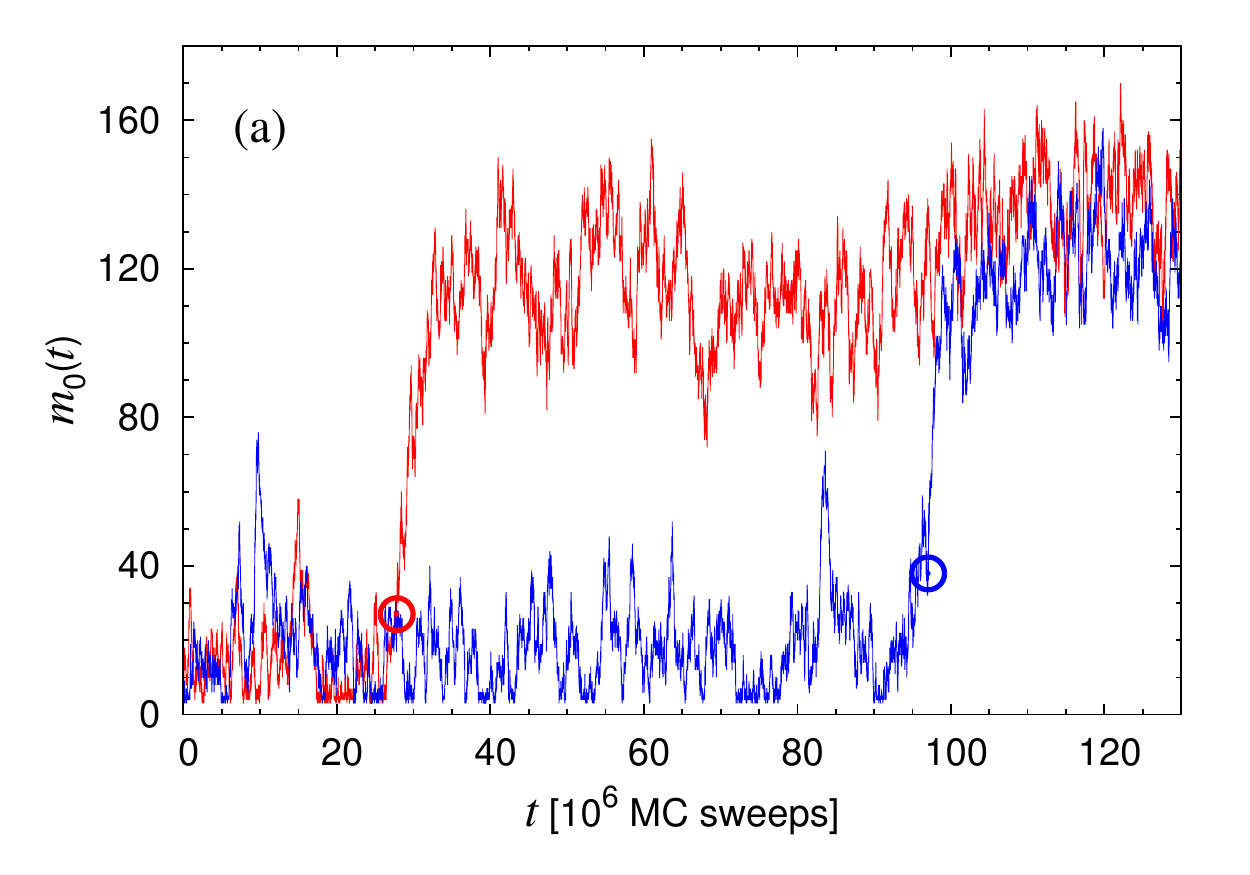}
\includegraphics[width=7.5cm]{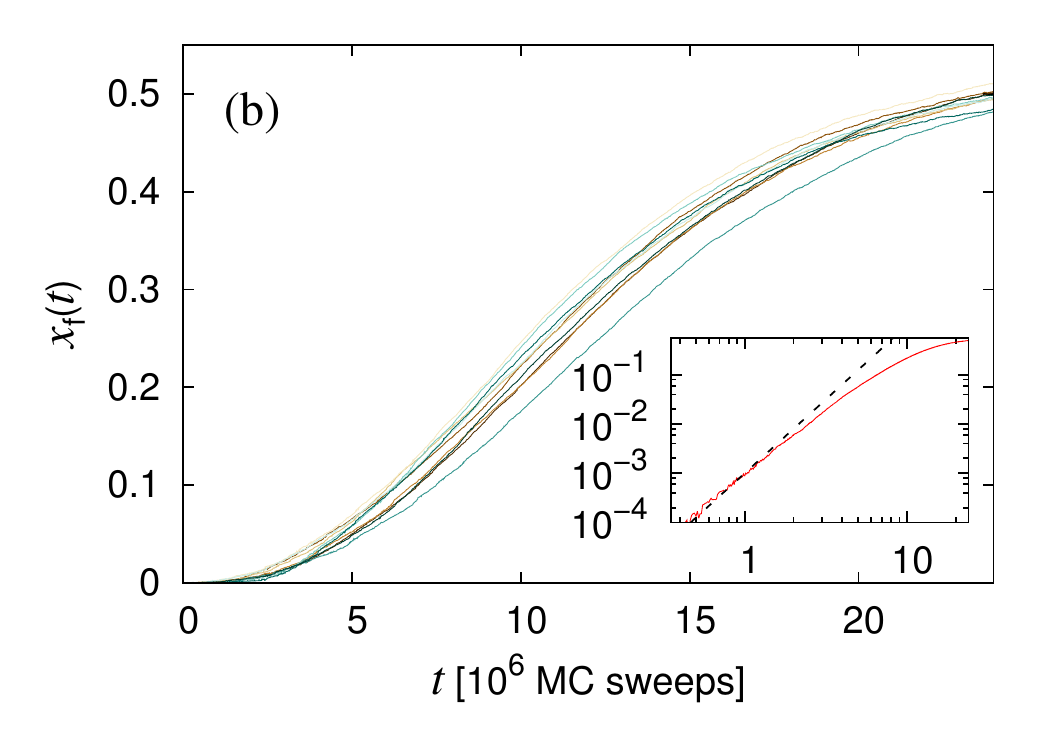}
\caption{Monte Carlo kinetics at $T=0.66$. (a) Mass of the largest 
aggregate, $\mo(t)$, against time, $t$, in two representative runs 
with $N=256$ and $L=64$. Circles indicate where the width of the
largest aggregate, $\wo$, switches from $\wo<3.5$ to $\wo>3.5$. Time is given
in sweeps, where one sweep consists of $N$ single-peptide updates.  
(b) Mass fraction of fibril-like aggregates, $\xf(t)$, 
in 10 independent runs with $N=131072$, $L=512$.
Inset: log-log plot of the average over the 10 runs vs. $t$. 
The dashed line corresponds to a cubic growth, $\xf(t)\propto t^3$.}  
\label{fig:kin}
\end{figure}

We first stick to the system size $N=256$ and $L=64$, which is useful 
for examining the formation of individual fibrils.   
Fig.~\ref{fig:kin}a shows the mass of the largest aggregate, 
$\mo(t)$, against Monte Carlo time, $t$, in two representative runs.
Both runs exhibit an apparent waiting phase, before a 
large aggregate suddenly appears. Unlike aggregates occurring in the 
initial phase, this large aggregate is stable to dissolution. Near the
jump in mass, a switch occurs in the width of the largest 
aggregate, $\wo$. With a tiny fraction of exceptions, $\wo$ is below 
3.5 before and above 3.5 after the switch point. Interestingly, as 
indicated in Fig.~\ref{fig:kin}a, this switch in width occurs immediately
before the sharp increase in mass. This suggests that the change in width
is a critical event that renders the aggregate growth-competent. 
This finding matches perfectly with the shape of the $p(l,w)$
distribution (Fig.~\ref{fig:lw}).   

The kinetics can also be studied for much larger systems,
which makes it possible to test in a direct manner whether or
not the model captures the sigmoidal behavior observed 
experimentally. Fig.~\ref{fig:kin}b shows the time evolution 
of the mass fraction of fibril-like aggregates, $\xf(t)$, in 10
independent runs for $N=131072$ and $L=512$ (same 
concentration as before).  
We define an aggregate as fibril-like 
if $w>3.5$, $w$ being the width, which ensures stability to
dissolution. 

Comparison of these 10 runs (Fig.~\ref{fig:kin}b) shows that, 
for this system size, the kinetics are indeed sigmoidal and  
highly reproducible, as observed in bulk 
experiments~\cite{Hellstrand:10}. At the end
of the runs, the simulation box typically contains between 40--50 
spontaneously formed fibrils
(see Supplemental Material, Fig.~S2~\cite{SUPP}), 
with an average length and width of 
$l\sim 210$  and $w\sim 7$, respectively. 
Inspection shows that
the nucleation of new fibrils stops after 
roughly $10^7$ Monte Carlo sweeps. Existing 
fibrils continue to grow beyond that point, but eventually $\xf(t)$  
levels off, due to monomer depletion. 

Our kinetic simulations, which do not include fragmentation 
events, may be compared to the classical Oosawa theory 
for homogeneous polymerization~\cite{Oosawa:62}. 
In particular, this theory predicts the initial growth to be
quadratic in time. Our data (Fig.~\ref{fig:kin}b) are well described 
by a cubic growth for small $t$, $\xf(t)\propto t^3$. 
That the exponent appears to be different than it is in the 
Oosowa theory is expected, because nucleation involves 
more than a single step in our model.    

In this Letter, we have presented a simple model for amyloid 
formation, where the nucleation of fibrils can be studied 
without any prior assumptions on the structure of the aggregates 
involved. The formation of aggregated structures with  
a few $\beta$-sheet layers has been observed in many previous
simulations, also at the atomic level~\cite{Li:08,Cheon:11}. 
Here, we have used systems much larger than in previous 
studies, to be able to examine the interplay 
between aggregate length and width in fibril nucleation.      
Our study shows that in this model the width of a growing 
aggregate plays a key role; to reach a given length, 
a minimum width is required, and to increase in width the 
aggregate has to overcome major free-energy barriers. Due to 
these barriers, the nucleation of a fibril occurs in distinct steps. 
The present study focused on the spontaneous aggregation 
of  free peptides, but  the model may also be useful for studying 
surface-catalyzed aggregation and the effects of a confining geometry. 

As in any model, simplicity is both a strength and a limitation.  The 
final width of a growing fibril depends most likely on geometric
factors left out in our model, such as twist. The question of what sets
the final width is therefore beyond the scope of the present work.  
The assumption that internal degrees of freedom 
can be integrated out may be a good approximation for small 
flexible peptides, but is clearly poorly justified for a folded protein 
that has to partially unfold before aggregation takes place.   
Our model further ignores any possible cooperativity of the 
interactions involved~\cite{Tsemekhman:07,Linse:11}.  In our 
model, aggregation is a highly cooperative process, although 
driven entirely by pairwise additive interactions. 

\begin{acknowledgments}
The simulations were performed at the LUNARC facility, Lund
University. 
\end{acknowledgments}

%merlin.mbs 2010-03-15 4.21a (PWD, AO, DPC)

\end{document}